\documentclass[conference]{IEEEtran}
\IEEEoverridecommandlockouts
\usepackage{cite}
\usepackage{amsmath,amssymb,amsfonts}
\usepackage{graphicx} 
\usepackage{textcomp}
\usepackage{xcolor}
\def\BibTeX{{\rm B\kern-.05em{\sc i\kern-.025em b}\kern-.08em
    T\kern-.1667em\lower.7ex\hbox{E}\kern-.125emX}}

\newcommand{\equalcontrib}{\textsuperscript{$^*$}}

\usepackage{float}
\usepackage{multirow} % For multi-row cells
\usepackage{array}    % For defining column styles
\usepackage{booktabs} % For better horizontal lines
\usepackage{afterpage}
\usepackage[hidelinks]{hyperref}

\usepackage{algorithm}
\usepackage{algpseudocode}
\usepackage{amsmath}
\usepackage{caption} % In the preamble
\usepackage{soul}

\hypersetup{
    colorlinks=true,
    linkcolor=blue,      % or black for print
    citecolor=green,     % dark green or black for a more subdued look
    urlcolor=black,       % dark red or a different shade of blue for distinction
    anchorcolor=blue     % color for anchor text
}

\begin{document}

\title{Quantum-Assisted Correlation Clustering\\
% {\footnotesize \textsuperscript{*}Note: Sub-titles are not captured in Xplore and should not be used}
\thanks{This work has been partially funded by the German Ministry for Education and Research (BMB+F) in the project QAIAC-QAI2C under grant 13N17167.}%

\thanks{$^*$ These authors contributed equally to this work.}
}

% \author{
%     \textsuperscript{2}Antonio Macaluso\textbf{$^*$} \IEEEcompsocitemizethanks{\IEEEcompsocthanksitem\IEEEauthorrefmark{1}\textbf{Equal contribution}},
%     \textsuperscript{1,2}Supreeth Mysore Venkatesh\textbf{$^*$},
%     \textsuperscript{3}Diego Arenas,\\
%     \textsuperscript{2}Matthias Klusch, 
%     \textsuperscript{1,3}Andreas Dengel\\
%     \textsuperscript{1}\textit{University of Kaiserslautern-Landau (RPTU)}, Kaiserslautern, Germany \\
%     \textsuperscript{2}\textit{German Research Center for Artificial Intelligence (DFKI)}, Saarbruecken, Germany \\
%     \textsuperscript{3}\textit{German Research Center for Artificial Intelligence (DFKI)}, Kaiserslautern, Germany \\
%     {\texttt{\{\href{mailto:antonio.macaluso@dfki.de}{antonio.macaluso}, \href{mailto:supreeth.mysore@dfki.de}{supreeth.mysore}, \href{mailto:diego.arenas@dfki.de}{diego.arenas},}} \\ \texttt{\href{mailto:matthias.klusch@dfki.de}{matthias.klusch}, \href{mailto:andreas.dengel@dfki.de}{andreas.dengel}\}@dfki.de}\\
%     {\texttt{\href{mailto:supreeth.mysore@rptu.de}{supreeth.mysore@rptu.de}}}
% }

\author{
    Antonio Macaluso\textsuperscript{1}\equalcontrib,
    Supreeth Mysore Venkatesh\textsuperscript{1,2}\equalcontrib,
    Diego Arenas\textsuperscript{3},\\
    Matthias Klusch\textsuperscript{1}, 
    Andreas Dengel\textsuperscript{1,3}\\
    \textsuperscript{1}\textit{German Research Center for Artificial Intelligence (DFKI)}, Saarbruecken, Germany \\
        \textsuperscript{2}\textit{University of Kaiserslautern-Landau (RPTU)}, Kaiserslautern, Germany \\
    \textsuperscript{3}\textit{German Research Center for Artificial Intelligence (DFKI)}, Kaiserslautern, Germany \\
    % \href{mailto:supreeth.mysore@rptu.de}
    \{\texttt{antonio.macaluso}, 
    % \href{mailto:antonio.macaluso@dfki.de}
    % \href{mailto:marlon.nuske@dfki.de}
    {\texttt{diego.arenas}},\\
    % \href{mailto:matthias.klusch@dfki.de}
    {\texttt{matthias.klusch}}, 
    % \href{mailto:andreas.dengel@dfki.de}
    {\texttt{andreas.dengel}}\}\texttt{@dfki.de}\\
    \texttt{supreeth.mysore@rptu.de}
}

% \author{
% \IEEEauthorblockN{Antonio Macaluso^*}
% \IEEEcompsocitemizethanks{\IEEEcompsocthanksitem\IEEEauthorrefmark{1}equal contribution}
% \IEEEauthorblockA{\textit{German Research Center for Artificial Intelligence} \\
% Saarbruecken, Germany \\
% antonio.macaluso@dfki.de}
% \and
% \IEEEauthorblockN{Supreeth Mysore Venkatesh^*}
% \IEEEauthorblockA{\textit{University of Kaiserslautern} \\
% Kaiserslautern, Germany \\
% supreeth.mysore@rptu.de}
% \IEEEauthorblockA{\textit{German Research Center for Artificial Intelligence} \\
% Saarbruecken, Germany \\
% supreeth.mysore@dfki.de}
% \and
% \IEEEauthorblockN{Diego Arenas}
% \IEEEauthorblockA{\textit{German Research Center for}\\ 
% \textit{Artificial Intelligence} \\
% Kaiserslautern, Germany \\
% diego.arenas@dfki.de}
% \and
% \IEEEauthorblockN{Matthias Klusch}
% \IEEEauthorblockA{\textit{German Research Center for}\\ 
% \textit{Artificial Intelligence} \\
% Saarbruecken, Germany \\
% matthias.klusch@dfki.de}
% \and
% \IEEEauthorblockN{Andreas Dengel}
% \IEEEauthorblockA{\textit{German Research Center for}\\ 
% \textit{Artificial Intelligence} \\
% Kaiserslautern, Germany \\
% andreas.dengel@dfki.de}
% }

\maketitle

% \begin{abstract}
% This document is a model and instructions for \LaTeX.
% This and the IEEEtran.cls file define the components of your paper [title, text, heads, etc.]. *CRITICAL: Do Not Use Symbols, Special Characters, Footnotes,
% or Math in Paper Title or Abstract.
% \end{abstract}

\begin{abstract}
This work introduces a hybrid quantum-classical method to correlation clustering, a graph-based unsupervised learning task that seeks to partition the nodes in a graph based on pairwise agreement and disagreement. In particular, we adapt GCS-Q \cite{venkatesh2023gcs}, a quantum-assisted solver originally designed for coalition structure generation, to maximize intra-cluster agreement in signed graphs through recursive divisive partitioning. The proposed method encodes each bipartitioning step as a quadratic unconstrained binary optimization problem, solved via quantum annealing. This integration of quantum optimization within a hierarchical clustering framework enables handling of graphs with arbitrary correlation structures, including negative edges, without relying on metric assumptions or a predefined number of clusters. 
Empirical evaluations on synthetic signed graphs and real-world hyperspectral imaging data demonstrate that, when adapted for correlation clustering, GCS-Q outperforms classical algorithms in robustness and clustering quality on real-world data and in scenarios with cluster size imbalance.
Our results highlight the promise of hybrid quantum-classical optimization for advancing scalable and structurally-aware clustering techniques in graph-based unsupervised learning.
\end{abstract}


% {\color{red}
% \section*{To-Do List}

% \begin{itemize}
%     \item Identify where generality is lost when moving from correlation-based to distance-based measures

%     \item Automatically determine the optimal number of clusters

%     \item Evaluate the quality of clustering results

%     \item Compare PAM vs. K-Means clustering algorithms

%     \item Use NMI to assess how well clustering aligns with the ground truth

%     \item Formalize the GCS-Q pseudo-core clustering approach

%     \item Mathematically define and relate correlation and distance

%     \item Compare graph clustering and correlation clustering

%     \item Analyze modularity (metric) in clustering

%     \item clusters with simular sizes, variance increseas gcs-q is better
% \end{itemize}
% }

% Expected experiments:

% \begin{itemize}
% \item Simulated data: Quality of the results

% \item for large size of cluster and training points
% \item different simulated data

% \end{itemize}

\section{Introduction}
Correlation clustering (CC) \cite{bansal2004correlation} is a graph-based paradigm that extends traditional clustering methods by directly modeling pairwise relationships. Unlike approaches such as \textit{k}-means or hierarchical clustering, which require data to be embedded in a metric space and rely on geometric proximity, CC operates on a weighted graph where edges represent real-valued affinities. Importantly, these weights can be negative, capturing dissimilarities and more general forms of conflict or disagreement between data points. As a result, it can naturally capture non-metric, asymmetric, or context-dependent relations between objects, making it more flexible in modeling diverse types of data, including those where distances are ill-defined or unavailable.
%Correlation clustering provides a versatile framework for partitioning data based on the sign and strength of pairwise relationships, making it particularly effective in settings where interactions can be both positive and negative. 
In social network analysis \cite{doreian1996partitioning}, CC enables the detection of communities by grouping individuals based on patterns of affinity or antagonism. In recommendation systems \cite{charikar2005clustering}, it supports collaborative filtering by accounting for both agreement and disagreement in user preferences. Applications also span natural language processing \cite{bansal2004correlation}, where CC helps group texts based on semantic relationships such as synonymy and antonymy, as well as in bioinformatics \cite{bhattacharya2008divisive}, where it has been used to identify groups of genes with similar patterns of variation in gene expression data, distinguishing co-expressed from anti-correlated gene profiles. Notably, CC plays a role in Earth Observation (EO), particularly in the analysis of high-dimensional hyperspectral imagery, where it helps identify clusters within subspaces that are not necessarily aligned with the coordinate axes \cite{isprs-annals-II-8-111-2014}. In this case, CC can detect redundant spectral bands, reducing dimensionality and improving interpretability without relying on strict axis-aligned projections. %This ability to work in arbitrarily oriented subspaces makes CC especially valuable for EO tasks.  

Classical methods for CC typically rely on heuristics and assumptions about the underlying data structure to achieve computational efficiency \cite{bansal2004correlation,ailon2009correlation}. These assumptions, such as spherical clusters in \textit{k}-means or connectivity in spectral clustering, guide the algorithm's behavior but may not hold in all real-world scenarios. As a result, there is often a trade-off between the quality of the clustering solution and the efficiency of the heuristic used to obtain it. While these approximations allow algorithms to scale to large datasets, they may also lead to suboptimal or biased partitions when the true data distribution deviates from the assumed model.

To overcome these limitations, it is useful to revisit CC from a more general, graph-theoretic perspective—one that naturally connects to the problem of coalition structure generation (CSG) in induced subgraph games (ISGs) \cite{deng1994complexity}. In ISGs, each node corresponds to an agent, and the value of a coalition is defined as the sum of the weights of the edges within the subgraph it induces. When correlation clustering is applied to a signed or weighted graph to maximize intra-cluster agreement—namely, the total weight of edges within clusters—the problem becomes equivalent to identifying an optimal coalition structure in an ISG. This formulation also aligns with the broader task of community detection\cite{zitnik2018prioritizing,avrachenkov2018network}. Both frameworks aim to partition the graph to maximize internal cohesion, and an optimal solution in one domain translates directly to the other.

% The graph-centric perspective in CC sets the stage for a natural connection to coalition structure generation (CSG) in induced subgraph games (ISGs)\cite{deng1994complexity}, a class of cooperative games where nodes (agents) form coalitions, and the value of a coalition is determined by the sum of the weights of the edges it induces. 
% In particular, when CC is formulated on a weighted graph with both positive and negative edge weights, and the objective is to maximize intra-cluster agreement, i.e., the total weight of edges within clusters, the problem becomes equivalent to CSG in an ISG \cite{zitnik2018prioritizing,avrachenkov2018network}. In this case, both frameworks aim to partition the graph in a way that maximizes the internal cohesion of each group, and the optimal solution to one is directly optimal for the other.

However, the equivalence holds only under specific conditions. 
CC becomes distinct from CSG in ISGs when the objective is to minimize disagreement rather than maximize agreement, when edge weights are limited to binary labels without magnitude, or when the context lacks game-theoretic considerations such as strategic behavior or fairness. 
In contrast, CSG in ISGs focuses on identifying partitions that maximize total value (social welfare) and negotiating stable and fair payoff allocations among rational agents. These stability considerations are formalized through solution concepts such as the Core, the Shapley value, and the Kernel~\cite{elkind2016cooperative}.
% In summary, CC provides a flexible and broadly applicable framework for graph partitioning, and under agreement-maximization with real-valued edge weights, it aligns formally with CSG in ISGs. However, its scope remains distinct when objectives shift or when strategic interactions among agents are not part of the problem model.

In this paper, we adapt GCS-Q \cite{venkatesh2023gcs}, a hybrid quantum annealing-based solver originally developed for CSG in ISGs, to address the task of correlation clustering on weighted graphs. By leveraging GCS-Q’s ability to maximize intra-cluster edge weights, we reinterpret the coalition value maximization objective as an agreement-based correlation clustering criterion. We benchmark the adapted algorithm against several classical clustering methods, including \textit{k}-means, %\remove{PAM} \new{Partitioning Around Medoids (PAM)}, 
Partitioning Around Medoids (PAM), Hierarchical, and Spectral clustering (for a comprehensive description of all clustering methods see \cite{kaufman2009finding}), across a wide range of settings. 
First, evaluating performance on simulated graph-based datasets with varying cluster size distributions to enable a rigorous assessment of each method’s ability to recover meaningful partitions, particularly in the presence of substantial imbalance. Cluster size variation is quantified using the Gini index, which captures configurations from nearly uniform to highly skewed. Complementary experiments on EO datasets are conducted to demonstrate the stability and effectiveness of the approach compared to classical methods on real-world data.
%First, evaluating performance on simulated graph-based datasets with varying group size distributions enables a rigorous assessment of each method’s ability to recover meaningful partitions, particularly under substantial imbalance. This variability is quantified using the Gini index, which captures configurations ranging from nearly uniform to highly skewed. Complementary experiments on real-world Earth Observation (EO) datasets further demonstrate that GCS-Q remains stable and effective, even in scenarios where classical methods tend to degrade.
%These additional results underscore the potential of the quantum-supported optimization routine embedded in GCS-Q for discrete clustering tasks framed as subgraph maximization problems, offering promising directions for hybrid quantum-classical approaches in graph-based unsupervised learning.

\section{Related Works}

The problem of Correlation Clustering focuses on partitioning a dataset based on pairwise relationships between data points, whether they are similar (positively correlated) or dissimilar (negatively correlated), without requiring an explicit geometric embedding or metric assumptions. This stands in contrast to traditional distance-based clustering methods, which rely on predefined distance functions and often fail to distinguish between different types of correlations \cite{thrun2021distance,ailon2009correlation,kaufman2009finding,murtagh2012algorithms}.
%\remove{For example, in Euclidean or cosine distance-based approaches, two points may be equidistant from a reference point while exhibiting opposite correlation patterns, a nuance that such methods inherently overlook.}
% \new{More importantly, the objective in distance-based approaches is to minimize the distance between the centroid and the cluster members, which is not the same as maximizing agreements.}
The CC problem is known to be NP-hard \cite{levorato2017evaluating}, and among classical approaches, spectral clustering \cite{von2007tutorial} has emerged as the most effective method for correlation-based tasks \cite{ding2024survey,7394552,mcsherry2001spectral}. By constructing a similarity graph and analyzing the spectral properties (eigenvalues and eigenvectors) of its Laplacian, spectral clustering captures the global structure of the data, including indirect relationships that are often critical in correlation graphs. This enables the identification of coherent groups of positively correlated points, even when they are not spatially adjacent in the original feature space.

A significant limitation shared by most classical clustering methods is the need to specify the number of clusters \(k\) in advance. This imposes a reliance on model selection heuristics or costly cross-validation procedures, where multiple values of \(k\) must be evaluated to identify the optimal partitioning. Such trial-and-error approaches hinder scalability and interpretability, especially in exploratory settings where the true number of clusters is unknown. %\new{Moreover, most classical algorithms tend to find clusters of equal size, which is not always the case.}

Recently, quantum approaches have been proposed to improve clustering by offloading specific computational routines to quantum hardware. In particular, gate-based quantum algorithms \cite{lloyd2013quantum,aimeur2013quantum,kerenidis2019q} offer theoretical speedups in fault-tolerant settings, but empirical evaluations are limited to very small datasets and depend heavily on classical pre- and post-processing. Furthermore, these methods do not introduce fundamentally new clustering strategies, but focus on accelerating existing algorithms. %Additionally, they do not inherently address key challenges such as automatically determining the number of clusters \(k\) or ensuring robustness to noisy or ambiguous correlation structures. 
Quantum annealing-based approaches to graph partitioning typically reformulate the entire clustering problem as a quadratic unconstrained binary optimization (QUBO) tailored to specific applications \cite{kalra2018portfolio}, but are not scalable and generalizable.

\section{Methodology}

Our proposed method follows the principles of hierarchical divisive clustering \cite{venkatesh2023gcs,murtagh2012algorithms}, a top-down approach that starts with a single cluster containing all nodes and recursively partitions it into smaller, more coherent clusters. Unlike agglomerative strategies that rely on local distances and pairwise merges, divisive methods operate by globally optimizing a splitting criterion at each step. This framework is particularly well-suited for correlation clustering, where the objective is to group similar nodes and separate dissimilar ones, based on edge weights representing pairwise agreement/disagreement.
The \textit{GCS-Q algorithm} \cite{venkatesh2023gcs}, originally developed for coalition structure generation in induced subgraph games, naturally fits this paradigm. We adapt it here for correlation clustering by reinterpreting its coalition value maximization 
objective as the task of maximizing \textit{intra-cluster agreement}. 

Given a weighted undirected graph \( G = (V, E, w) \), where \( w_{ij} \in \mathbb{R} \) reflects the affinity between nodes \( i \) and \( j \), the goal is to find a partition \( \mathcal{C} = \{C_1, \dots, C_k\} \) that maximizes:
\begin{equation}
\max \sum_{i,j \in S} w_{ij} \cdot \mathbb{I}[x_i = x_j].
\label{eq:max-agree}
\end{equation}
Here, \( x_i \in \{0,1\} \) is a binary variable indicating the cluster assignment of node \( i \). For a weighted graph with both positive and negative edge weights \( w_{ij} \), Eq.~\eqref{eq:max-agree} corresponds to the \textit{maximize-agreement} variant of CC and is equivalent to minimizing the total weight of edges cut across clusters. Therefore, the goal is to maximize intra-cluster agreement, where positively weighted edges connect nodes within the same cluster, and negatively weighted edges span different clusters.
To formalize this, we note that the indicator function for agreement between node labels can be expressed as:
\begin{equation}
\mathbb{I}[x_i = x_j] = 1 - (x_i + x_j - 2x_i x_j),    
\end{equation}
which evaluates to 1 when \( x_i = x_j \), and 0 otherwise. Using this identity, the agreement maximization objective becomes:
\begin{equation}
\max_{x \in \{0,1\}^n} \sum_{i < j} w_{ij} (1 - x_i - x_j + 2 x_i x_j).    
\end{equation}

Dropping the constant term (which does not affect the optimization), this formulation is fully compatible with QUBO solvers and equivalent in objective value to the disagreement minimization version used in GCS-Q. It provides a clear and implementable pathway for optimization on quantum annealing to perform correlation clustering based on agreement maximization.

Importantly, a key distinction between GCS-Q \cite{venkatesh2023gcs} and classical hierarchical divisive clustering lies in the splitting strategy. Traditional divisive methods typically rely on heuristic or local distance-based criteria (e.g., edge betweenness or spectral cuts), which may fail to identify globally optimal partitions. Moreover, these classical approaches generally assume metric properties and non-negative similarities, making them ill-suited for settings involving signed relationships or negative correlations. In contrast, the adaptation of GCS-Q inherits the advantages of the original formulation \cite{venkatesh2023gcs} and explicitly optimizes a global objective over the current subgraph at each iteration using combinatorial search guided by quantum annealing. This enables GCS-Q to evaluate all possible binary bipartitions simultaneously, thereby avoiding the suboptimal greedy decisions often made by classical methods. As a result, GCS-Q is expected to outperform classical divisive approaches, particularly when applied to graphs with both positive and negative edge weights. Moreover, since GCS-Q does not rely on an explicit notion of distance, it is also expected to outperform its classical counterpart when clustering is driven by general correlation structures rather than spatial proximity.

The complete procedure is summarized in Algorithm \ref{alg:GCS-Q}, where the original game-theoretic stopping criterion in \cite{venkatesh2023gcs} is replaced by one based on intra-cluster agreement. %This adaptation enables GCS-Q to serve as a principled and effective solver for correlation clustering, while taking advantage of quantum-supported combinatorial optimization.

% \begin{algorithm}
% %\caption{GCS-Q for Correlation Clustering}
% \caption{GCS-Q for Correlation Clustering}
% \begin{algorithmic}[1]
% \Require Weighted undirected graph $G = (V, E, w)$
% \Ensure Partition $\mathcal{C} = \{C_1, C_2, \dots, C_k\}$ maximizing intra-cluster agreement

% \State $\mathcal{C} \gets \emptyset$
% \State $\text{Queue} \gets \{V\}$ \Comment{Initialize with the full node set}

% \While{Queue $\neq \emptyset$}
%     \State $S \gets \text{Queue.pop}()$
%     \State Construct subgraph $G_S = (S, E_S, w_S)$
%     \State Formulate QUBO for max $\sum_{i,j \in S} w_{ij} \cdot \mathbb{I}[x_i = x_j]$
%     \State Solve QUBO on quantum annealer to obtain binary solution $x$
%     \State Let $C \gets \{i \in S \mid x_i = 1\}$, $\bar{C} \gets S \setminus C$
    
%     \If{$\text{cut}(C, \bar{C}) \leq 0$ \textbf{or} $C = \emptyset$ \textbf{or} $\bar{C} = \emptyset$}
%         \State $\mathcal{C} \gets \mathcal{C} \cup \{S\}$ \Comment{No beneficial split}
%     \Else
%         \State $\text{Queue} \gets \text{Queue} \cup \{C, \bar{C}\}$
%     \EndIf
% \EndWhile

% \State \Return $\mathcal{C}$
% \end{algorithmic}
% \caption{GCS-Q}
% \label{alg:GCS-Q}
% \end{algorithm}
\begin{algorithm}
\caption{GCS-Q for Correlation Clustering}
\begin{algorithmic}[1]
\Require Weighted graph $G = (V, E, w)$
\Ensure Clustering $\mathcal{C} = \{C_1, C_2, \dots, C_k\}$

\State $\mathcal{C} \gets \emptyset$, \quad $\text{Queue} \gets \{V\}$
\While{Queue is not empty}
    \State $S \gets \text{Queue.pop}()$, \quad $G_S \gets$ subgraph of $S$
    \State Solve QUBO: $\max \sum_{i,j \in S} w_{ij} \cdot \mathbb{I}[x_i = x_j]$
    \State $C \gets \{i \in S \mid x_i = 1\}$, \quad $\bar{C} \gets S \setminus C$
    \If{$\text{cut}(C, \bar{C}) \leq 0$ or $C = \emptyset$ or $\bar{C} = \emptyset$}
        \State $\mathcal{C} \gets \mathcal{C} \cup \{S\}$
    \Else
        \State $\text{Queue} \gets \text{Queue} \cup \{C, \bar{C}\}$
    \EndIf
\EndWhile
\State \Return $\mathcal{C}$
\end{algorithmic}
\label{alg:GCS-Q}
\end{algorithm}

% \begin{algorithm}
% \caption{GCS-Q: Quantum-Assisted Correlation Clustering}
% \begin{algorithmic}[1]
% \Require Weighted undirected graph $G = (V, E, w)$ representing pairwise similarities and dissimilarities
% \Ensure Clustering $\mathcal{C} = \{C_1, C_2, \dots, C_k\}$ that maximizes intra-cluster agreement

% \State Initialize clustering set: $\mathcal{C} \gets \emptyset$
% \State Initialize queue with the full node set: $\text{Queue} \gets \{V\}$

% \While{Queue is not empty}
%     \State Extract current cluster candidate: $S \gets \text{Queue.pop}()$
%     \State Construct subgraph $G_S = (S, E_S, w_S)$ induced by $S$
    
%     \State Encode bipartitioning objective as a QUBO:
%     \[
%     \max \sum_{i,j \in S} w_{ij} \cdot \mathbb{I}[x_i = x_j]
%     \]
%     \State Solve QUBO using a quantum annealer to obtain cluster assignment $x$
%     \State Define two subclusters: 
%     \[
%     C \gets \{i \in S \mid x_i = 1\}, \quad \bar{C} \gets S \setminus C
%     \]

%     \If{split is not beneficial (i.e., $\text{cut}(C, \bar{C}) \leq 0$) or $C = \emptyset$ or $\bar{C} = \emptyset$}
%         \State Add $S$ as a final cluster: $\mathcal{C} \gets \mathcal{C} \cup \{S\}$
%     \Else
%         \State Enqueue the resulting subclusters: $\text{Queue} \gets \text{Queue} \cup \{C, \bar{C}\}$
%     \EndIf
% \EndWhile

% \State \Return Final clustering $\mathcal{C}$
% \end{algorithmic}
% \label{alg:GCS-Q}
% \end{algorithm}

\section{Experiments}

To evaluate the effectiveness and robustness of our proposed GCS-Q-based correlation clustering approach, we conduct a comprehensive set of experiments on both synthetic and real-world datasets. The experimental design aims to assess performance relative to classical methods under diverse structural conditions, including varying cluster size distributions and real-world configurations, using both internal (modularity) and external (Normalized Mutual Information) evaluation metrics where applicable.

\begin{figure*}[htbp]
  \centering
  \includegraphics[width=\textwidth]{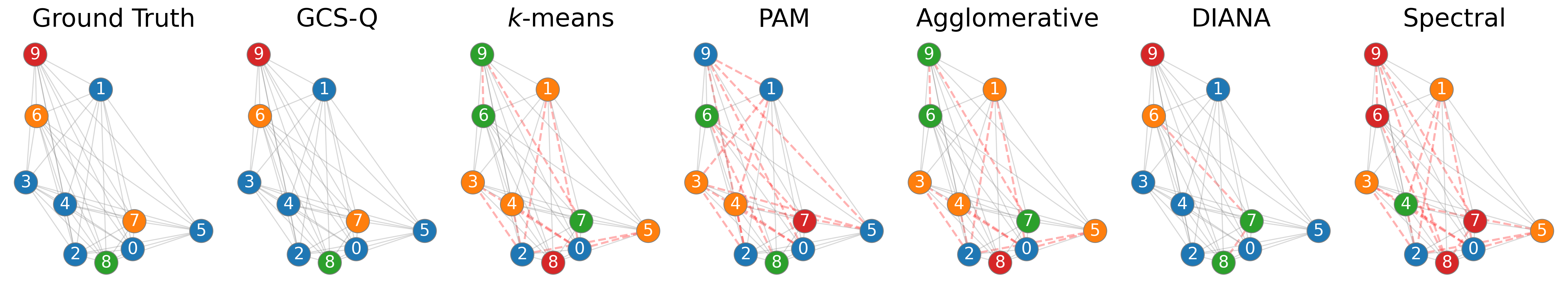}
  % \vspace*{-16pt}
  \caption{The ground truth clusters and the outputs of various algorithms for a balanced graph with 10 nodes and edge weights in the range $[-1, 1]$ representing correlations. Red dashed edges highlight inconsistent assignments: a red edge between nodes in the same cluster indicates a negative correlation, while a red edge between nodes in different clusters indicates a positive correlation. In other words, red dashed lines mark violations of clustering agreement. All other edges are shown in grey, either because they are neutral or consistent with the clustering. These inconsistencies are usually minimized by GCS-Q and are more frequent in classical methods.
 }
  %Ground truth clusters and the outputs of various algorithms for a balanced graph with 10 nodes, where edge weights in the range $[-1, 1]$ represent pairwise correlations. Red dotted edges indicate negative correlations between nodes assigned to the same cluster, while solid grey edges connect either positively correlated nodes within a cluster or nodes from different clusters. Notably, several classical methods (e.g., PAM, DIANA, Spectral) produce clusters that include negatively correlated nodes---as evidenced by the presence of red edges within clusters---highlighting a limitation in their ability to handle signed relationships.
  \vspace*{-16pt}
  \label{fig:clustering_methods_example}
\end{figure*}

\subsection{Settings}

\subsubsection{Datasets}

In this work, we evaluate clustering performance using two types of datasets: synthetically generated signed graphs with known clustering labels and four real-world hyperspectral image datasets.

\paragraph{Synthetic Signed Graph Dataset} 

\begin{figure}[H]
    \centering
    \includegraphics[width=\linewidth]{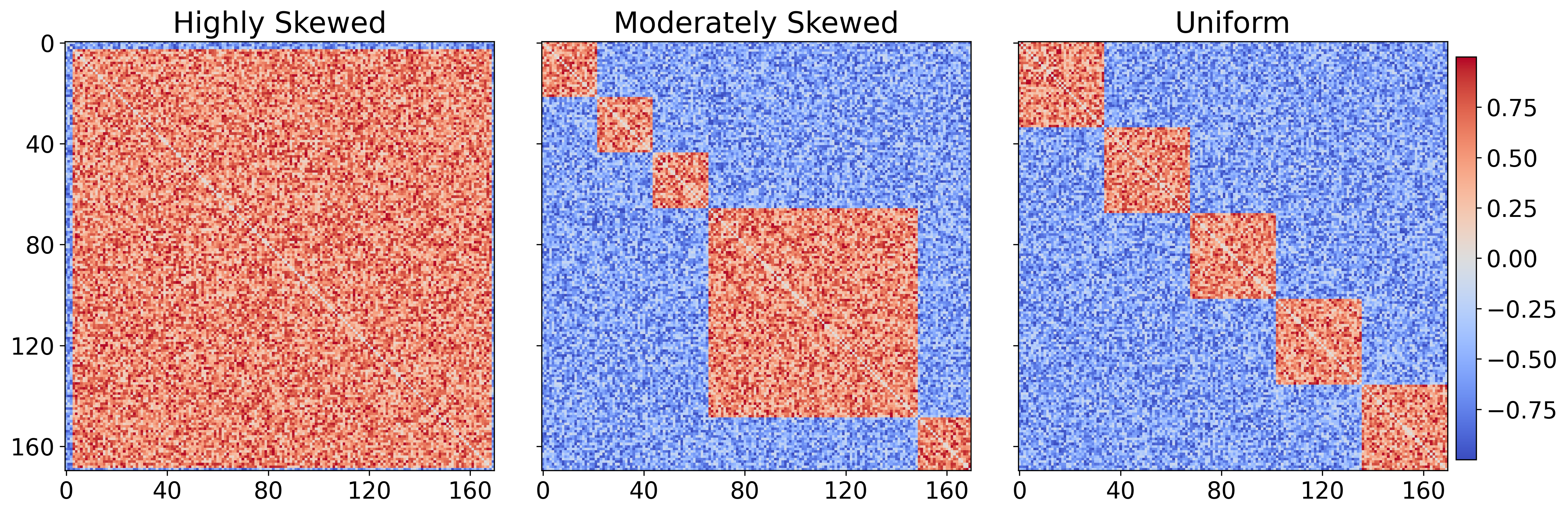}
        \caption{Correlation matrices of three synthetic signed graphs used in our experiments. The examples correspond to a setting with $5$ clusters. Warmer colors indicate strong positive correlations (intra-cluster), while cooler tones represent negative correlations (inter-cluster). The corresponding Gini index and Cluster Size Ratio for each matrix are reported in Table~\ref{tab:Gini}.
}
    \label{fig:data}
\end{figure}

We generated three classes of synthetic signed graphs varying structural and distributional conditions. Each class is represented as a weighted adjacency matrix (Fig.~\ref{fig:data}), where edge weights denote pairwise correlation coefficients between nodes. The graphs differ in the clarity of their block structure, reflecting the degree of separation and the relative sizes of the underlying clusters.
To simulate varying levels of cluster size disparity, we systematically adjusted the Gini index across the three types of datasets. The Gini index, ranging from $0$ (perfectly uniform cluster sizes) to $1$ (complete inequality), provides a quantitative measure of imbalance in cluster membership. This variation enables a controlled assessment of how different clustering algorithms respond to increasingly skewed distributions.
All generated graphs are balanced \cite{aref2019balance} in the sense that their vertices can be divided into two clusters such that all intra-cluster edges are positive and all inter-cluster edges are negative. Our synthetic data generalizes this notion to multiple clusters, maintaining internal coherence through positive correlations and ensuring negative separation across clusters. This guarantees that clusters are well-separated, allowing for a clean evaluation of clustering performance under ideal yet structurally diverse conditions.
The three matrices shown in Fig.~\ref{fig:data} are examples of generated graphs with 170 nodes, which represents the practical upper limit for fully connected graphs that can currently be embedded on available quantum annealing hardware. However, it is important to note that assuming a certain degree of sparsity in the graph structure allows QUBO formulations to scale to larger problem instances \cite{venkatesh2024q}.

\begin{table}[h]
\centering
% \caption*{\textbf{TABLE I}}
\vspace{1em}
\begin{tabular}{lccc}
\hline
\textbf{Metric} & \textbf{High Skewed} & \textbf{Moderately Skewed} & \textbf{Uniform} \\
\hline
Gini Index                & 0.05  & 0.70  & 0.80  \\
Cluster Size Ratio           & 166.00 & 3.95  & 1.00  \\
\hline
\end{tabular}
\caption{Gini index and cluster size ratio (largest cluster size divided by smallest) for the three synthetic graphs in Fig.~\ref{fig:data}. The low Gini index case shows extreme cluster size disparity, with most nodes concentrated in a single cluster.}
\vspace{-1em}
\label{tab:Gini}
\end{table}

\paragraph{Hyperspectral Image Dataset}

We also test our methods on real-world data drawn from publicly available hyperspectral remote sensing scenes provided by the Universidad del País Vasco repository\footnote{\url{https://www.ehu.eus/ccwintco/index.php?title=Hyperspectral_Remote_Sensing_Scenes}}.
In this experiment, we perform correlation clustering on hyperspectral datasets (Indian Pines, Salinas, Pavia University, and Kennedy Space Center (KSC)) to identify redundant spectral bands \cite{isprs-annals-II-8-111-2014,rs14051156}. The data consists of hundreds of narrow, continuous spectral bands capturing detailed information across the electromagnetic spectrum. We calculate the Pearson correlation coefficients between bands, forming a correlation matrix, and use this matrix to perform clustering, grouping bands with high correlation to reduce data redundancy for further analysis \cite{Ghorbanian03102018,7394552}.

\subsubsection{Methods}
To evaluate the performance of GCS-Q, we compare several representative clustering methods. \textit{k}-means is a classical distance-based algorithm that partitions data by minimizing intra-cluster variance, assuming a Euclidean embedding. PAM is a more robust alternative to \textit{k}-means that selects actual data points as cluster centers, making it less sensitive to outliers and more effective for non-spherical clusters. Spectral clustering leverages the eigenstructure of a similarity graph to detect globally coherent clusters and is particularly suited for correlation-based tasks. Hierarchical clustering builds a nested tree of clusters based on pairwise similarities; in our study, we employ both agglomerative (bottom-up) and divisive (top-down, DIANA) variants to capture different structural perspectives. %, allowing multi-resolution analysis without requiring an initial number of clusters.

\subsubsection{Evaluation Metrics}

To assess the quality of the clustering results, we employ two complementary metrics aiming at capturing different aspects of clustering performance and providing a more comprehensive evaluation.

\paragraph{Normalized Mutual Information (NMI)} This is an external validation metric that quantifies the similarity between the clustering output and a known ground-truth partition. It measures the mutual dependence between the predicted labels and the true labels. NMI is normalized to lie in the range \([0, 1]\), where $0$ indicates no mutual information (i.e., independent assignments), and $1$ indicates a perfect match between the cluster assignments and the ground truth.

\paragraph{Modularity}
In practical situations, clustering algorithms are often applied to data for which the true cluster labels are unknown. In this context, the modularity metric \cite{newman2004finding} is essential for assessing the quality of clustering by comparing the observed fraction of intra-cluster edge weights to the expected fraction under a null model of random edge assignment.
 Given a division of a graph into clusters, modularity is defined as
\begin{equation}
Q = \sum_{i} (e_{ii} - a_i^2),    
\end{equation}
where \( e_{ii} \) is the fraction of edge weight connecting nodes within cluster \( i \), and \( a_i \) is the fraction of all edge weights connected to nodes in cluster \( i \). A higher value of \( Q \) indicates that the clustering captures significant structure in the graph, i.e., nodes are more positively connected within clusters than between them. In correlation clustering, this helps assess whether the clustering correctly groups positively correlated points while separating negatively correlated ones.

%number cluster given
%remove equation 4 
% captions

\subsection{Results}

\begin{figure}[h]
    \centering
    \includegraphics[width=\linewidth]{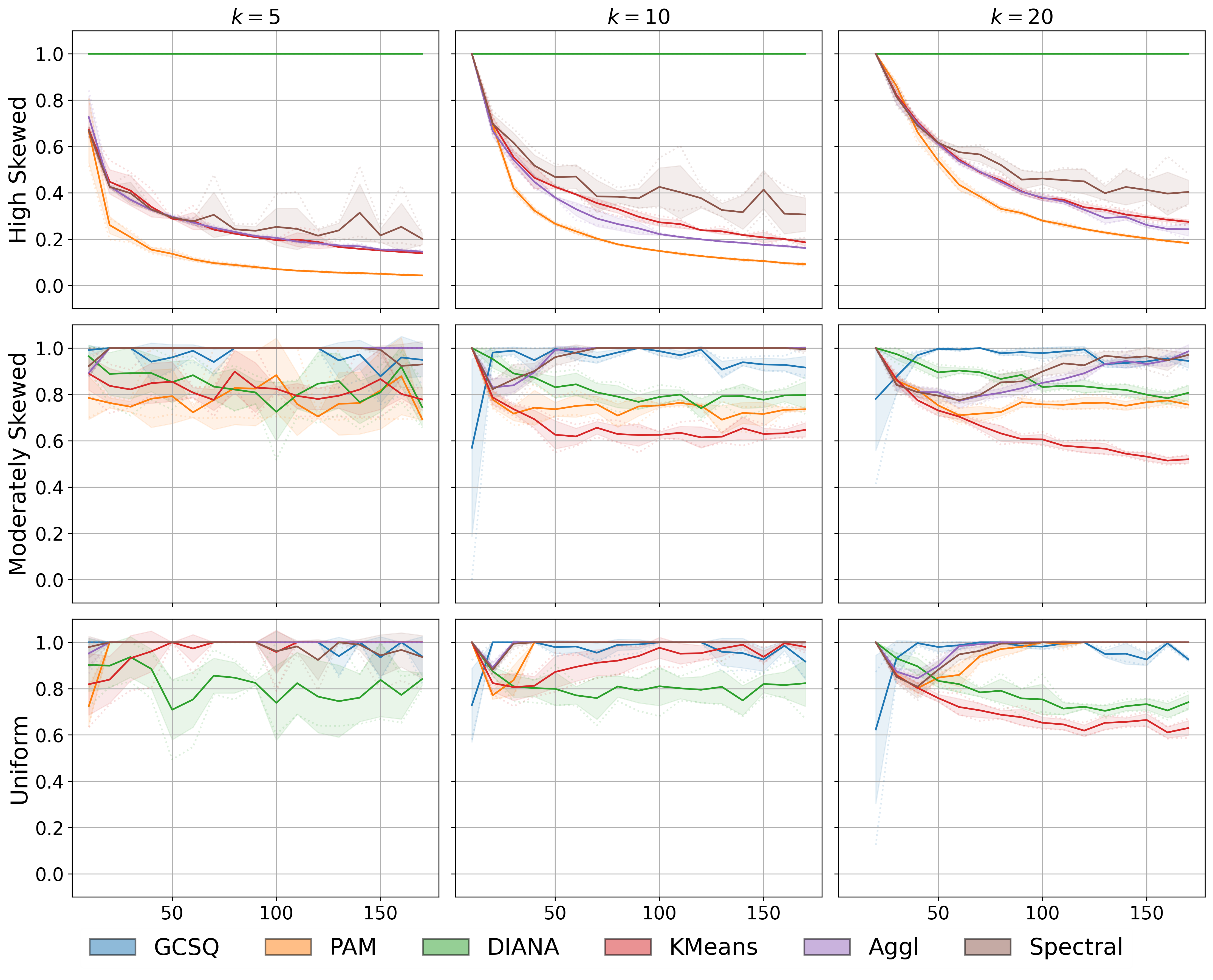}
    \caption{NMI scores across synthetic signed graphs with increasing cluster size disparity. %Each row corresponds to a different Gini index, representing increasing inequality in cluster sizes. Each column represents a different number of generated clusters, and each point within a plot denotes the result for a different total number of nodes. %GCS-Q consistently outperforms classical methods, particularly under high Gini conditions.
    }
    \label{fig:nmi_vs_gini}
\end{figure}

Figure~\ref{fig:nmi_vs_gini} illustrates the performance of several clustering algorithms in terms of NMI across different distributional settings---namely, highly skewed, moderately skewed, and uniform distributions---and for varying cluster counts ($k = 5, 10, 20$). 
Importantly, we provide the correct number of clusters to the classical methods to ensure a fair comparison, even though this information is typically unavailable in real-world scenarios.

Among the evaluated methods, GCS-Q consistently achieves strong performance, maintaining high NMI scores across all scenarios. While minor drops are observed in moderately skewed and uniform cases, GCS-Q remains competitive and exhibits remarkable stability compared to classical alternatives.
In highly skewed scenarios, DIANA and GCS-Q produce comparable results, which is expected: when $k - 1$ clusters contain singletons, DIANA’s greedy strategy of isolating one point per iteration aligns well with the underlying structure. However, in all other conditions, GCS-Q clearly outperforms DIANA, owing to its more comprehensive strategy that evaluates all possible bipartitions at each recursive step, as opposed to DIANA’s limited single-point exclusions.

Spectral clustering, though a state-of-the-art method for correlation clustering under balanced conditions, struggles with strong cluster size imbalances. In contrast, only GCS-Q and DIANA exhibit resilience under such distortions. Interestingly, in moderately skewed distributions with small $k$ (e.g., $k = 5, 10$), spectral clustering performs well and often matches GCS-Q’s results. However, as $k$ increases (e.g., $k = 20$), GCS-Q decisively outperforms all classical baselines, including spectral clustering.
Finally, in uniform distributions, GCS-Q continues to perform competitively. Although other methods improve in this more balanced setting, GCS-Q remains one of the most robust performers across all configurations, reinforcing its versatility and generalization capabilities.

Overall, these results confirm that GCS-Q offers a robust and high-performing alternative to classical clustering algorithms, particularly in challenging tasks involving signed graphs with significant cluster size disparity. Importantly, GCS-Q does not require the number of clusters $k$ to be provided in advance, as it automatically stops when no further bipartition increases intra-cluster agreement.
 Its ability to maintain high NMI across diverse structural conditions underscores its suitability for practical applications involving complex graph data.

To evaluate the performance of clustering methods in real-world scenarios, we apply correlation clustering to four hyperspectral remote sensing datasets—Indian Pines, Salinas, Pavia University, and KSC—sourced from the Universidad del País Vasco repository.
A preliminary analysis was performed to estimate the appropriate number of clusters for the classical methods. Specifically, the spectral gap criterion \cite{zelnik2004self} was used for spectral clustering, while the silhouette score \cite{rousseeuw1987silhouettes} was applied to the other methods.

%\begin{figure*}[htbp]
%  \centering
  %\includegraphics[width=\textwidth]{figures/finding-optimal-k-EO.png}
  % \vspace*{-16pt}
  %\caption{Silhouette scores for varying number of clusters ($k$) using k-means in the backend for the different hyperspectral datasets. The optimal $k$ value has been highlighted by the red vertical line.
 %}
  %\vspace*{-16pt}
  %\label{fig:finding-optimal-k-EO}
%\end{figure*}

\begin{figure}[h]
    \centering
    \includegraphics[width=\linewidth]{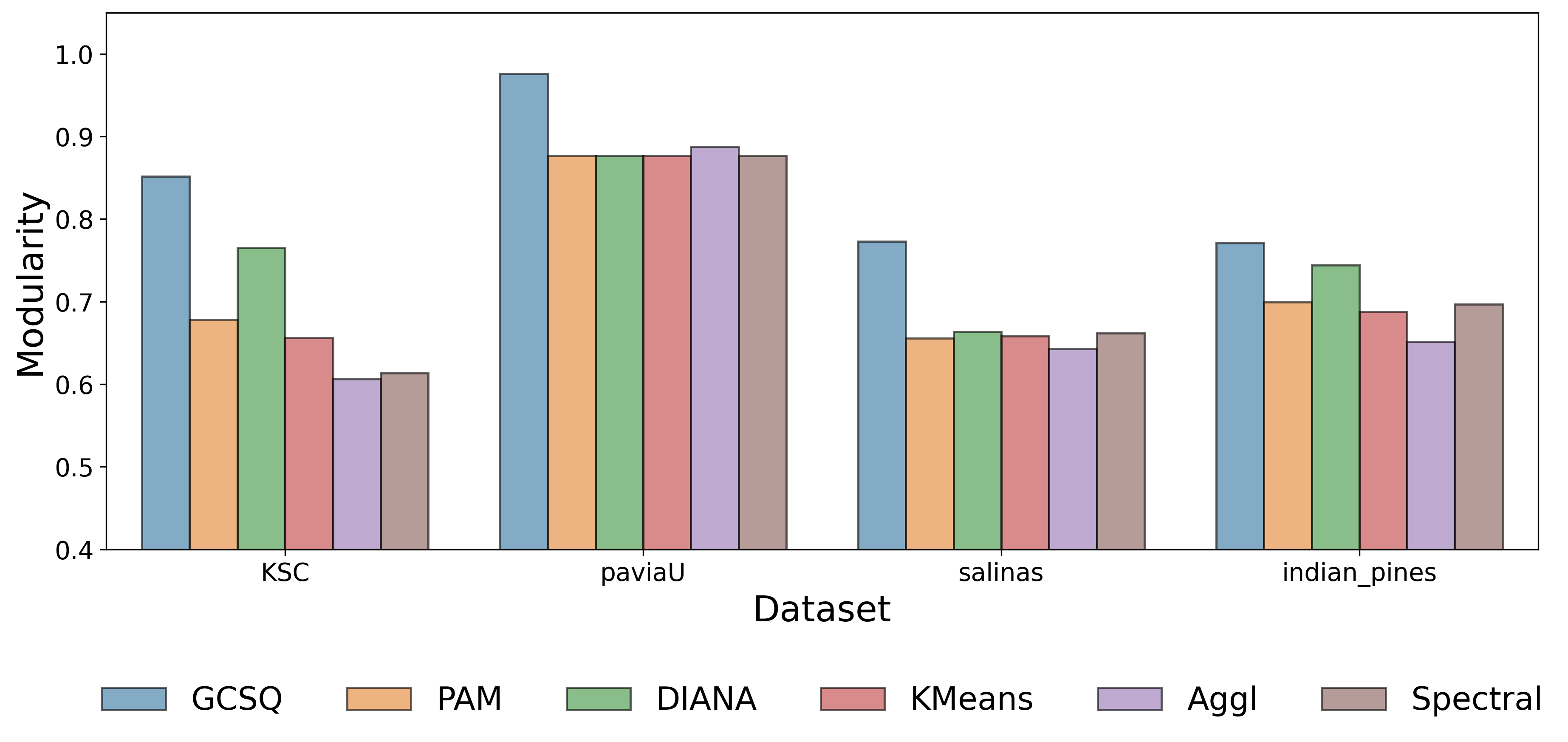}
\caption{Modularity scores for clustering hyperspectral bands across four real-world remote sensing datasets. GCS-Q consistently achieves superior modularity, indicating more coherent grouping of highly correlated spectral bands.}
    \label{fig:modularity}
\end{figure}

Figure~\ref{fig:modularity} presents the modularity scores achieved by each method across the four hyperspectral datasets. 
GCS-Q consistently obtains the highest modularity values, demonstrating its ability to identify strongly coherent spectral clusters. 
This advantage is particularly pronounced on the KSC, PaviaU and Salinas datasets, which are characterized by higher spectral complexity. 
Among classical methods, DIANA performs competitively and shows stable performance across all datasets, occasionally approaching GCS-Q's performance. 
In contrast, other classical approaches such as PAM, \textit{k}-means, and agglomerative hierarchical clustering tend to yield lower modularity scores, indicating less effective partitioning of the spectral correlation structure. 
%Spectral clustering performs reasonably well, especially on KSC and Indian Pines, but still generally lags behind GCS-Q. 

\section{Conclusion}
In this work, we proposed a novel adaptation of GCS-Q, a quantum-assisted solver originally developed for coalition structure generation, to address the problem of correlation clustering on signed graphs. By formulating the clustering task as a maximize-agreement objective over a weighted graph, we leveraged the power of quantum annealing to solve a sequence of QUBO problems that recursively identify meaningful partitions. Our method follows a divisive hierarchical strategy, avoiding the need to predefine the number of clusters and improving global optimization at each step.

Through extensive experiments on both synthetic signed graphs and real hyperspectral datasets, we demonstrated the robustness and superior clustering quality of GCS-Q, especially under challenging conditions such as high cluster size disparity. On synthetic data, GCS-Q achieved competitive NMI scores across varying Gini index levels, demonstrating superior performance and robustness to skewed cluster size distributions.
On real hyperspectral EO data, GCS-Q consistently outperformed classical baselines in terms of modularity, indicating better separation of correlated spectral bands.

These results confirm the viability of hybrid quantum-classical optimization for graph-based optimization, even when using noisy quantum annealing devices such as the D-Wave Advantage employed in this study. This encourages further exploration of quantum-assisted algorithms for unsupervised learning tasks. Future work includes expanding the experimental evaluation of GCS-Q, exploring alternative gate-based quantum solvers \cite{QuACS, qubit_efficient_encoding}, and applying the method to moderately sparse, large-scale real-world graph data beyond the remote sensing domain.

\bibliographystyle{IEEEtran}
\bibliography{references}

\end{document}